\documentclass[prx,a4paper,aps,onecolumn,superscriptaddress,11pt,longbibliography,accepted=2018-02-21]{quantumarticle}

\pdfoutput=1

\usepackage[utf8]{inputenc}
\usepackage[english]{babel}
\usepackage[T1]{fontenc}
\usepackage{amsmath}
\usepackage{hyperref}
\usepackage{tabularx}
\usepackage{tikz}

\usepackage[colorlinks=true]{hyperref}
 \newcommand{\ket}[1]{\ensuremath{\vert#1\rangle}}
 \newcommand{\bra}[1]{\ensuremath{\langle #1\vert}}

\newcommand{\kb}[2]{\ensuremath{\vert #1 \rangle \langle #2 \vert}}

\newtheorem{defin}{Definition}

\usepackage{array}
\newcolumntype{L}{>{\arraybackslash}m{8cm}}
\newcommand*{\etc}[1]{\textcolor{black}{#1}}

 \def\id{{\mathchoice {\rm 1\mskip-4mu l} {\rm 1\mskip-4mu l} {\rm 1\mskip-4.5mu l} {\rm 1\mskip-5mu l}}}

\begin{document}

\title{Magic state parity-checker \\ with pre-distilled components}
 \date{\today}
\author{Earl T. Campbell}
 \affiliation{Department of Physics \& Astronomy, University of Sheffield, Sheffield, S3 7RH, United Kingdom.}
 \email{earltcampbell@gmail.com}
 \homepage{https://earltcampbell.com/}
 \author{Mark Howard}
 \affiliation{Department of Physics \& Astronomy, University of Sheffield, Sheffield, S3 7RH, United Kingdom.}
 \email{m.howard@sheffield.ac.uk}
  \homepage{https://markhoward.info}
 
 % \maketitle
 
\begin{abstract}
Magic states are eigenstates of non-Pauli operators. One way of suppressing errors present in magic states is to perform parity measurements in their non-Pauli eigenbasis and postselect on even parity. Here we develop new protocols based on non-Pauli parity checking, where the measurements are implemented with the aid of pre-distilled multiqubit resource states. This leads to a two step process: pre-distillation of multiqubit resource states, followed by implementation of the parity check.  These protocols can prepare single-qubit magic states that enable direct injection of single-qubit axial rotations without subsequent gate-synthesis and its associated overhead. We show our protocols are more efficient than all previous comparable protocols with quadratic error reduction, including the protocols of Bravyi and Haah.
\end{abstract}

Error corrected quantum computers require additional gadgets and tricks to enable fully universal, fault-tolerant quantum computation (see Ref.~\cite{ReviewPaper} for a review).  Of the various competing approaches, magic state distillation is especially efficient.  At first, magic state distillation was conceived of as a way to implement the $T$ gate (also called the $\pi/8$ phase gate), with successive generations of protocols making ever greater improvements~\cite{BraKit05,Meier13,Bravyi12,Jones13,haah17magic}.  Any desired unitaries could be approximated by efficient synthesis of $T$ gates and Clifford gates, with recent years also bringing new, optimised synthesis methods~\cite{kliuchnikov13,gosset14,RS14,paetznick14,bocharov15}.  However, we can circumvent the need for synthesis if we instead prepare magic states tailored for injecting specific gates.  Preparing tailored single qubit magic states can directly provide single qubit rotations other than the $T$-gate and distillation routines for these states have been proposed~\cite{landahl13,jones13fourier,duclos15,smallAngle,haah17towers}.  Complex multi-qubit circuits can also be directly injected once multi-qubit magic states have been distilled~\cite{jones13b,eastin13,campbell17,campbell17b,haah17magic}.  

Here we propose a family of two-step protocols for distilling magic states that are tailored for injecting a desired single-qubit $Z$-axis rotation.  A distinctive feature is that the first step creates a multi-qubit magic state using the synthillation protocol~\cite{campbell17,campbell17b}.  In the second step, this multi-qubit resource is then used to fault-tolerantly perform a parity check in a non-Pauli basis.   Combined, the protocol takes in single qubit states and outputs single qubit states, with multi-qubit magic states appearing only fleetingly.  

A single round of our protocol will quadratically reduce errors using fewer inputs per output than any previous protocol with quadratic error reduction.  Higher order error reductions can be achieved by concatenation of our protocols.  Higher order reductions in noise are possible without concatenation by using sophisticated protocols that give a large error reduction in a single round~\cite{Jones13,haah17magic,haah2017examples,hastings2017sublog}, but there are important practical considerations for why one might favour a concatenated approach (see Sec.~\ref{sec_quadratic} for a discussion).

We begin by covering some basic notation.  Sec.~\ref{Overview} gives an overview of  our approach.  Note that the first step is our previously proposed synthillation method~\cite{campbell17,campbell17b}, so the details will not be repeated here.  Rather we focus on how non-Pauli parity checking is possible given these pre-distilled resources, giving a detailed explanation in Sec.~\ref{Sec_two_step}.   The protocol's performance is analysed in Sec.~\ref{Sec::noise}.  \etc{In Sec.~\ref{Sec_No_tri}, we present a small bonus result that proves equivalent performance is unlikely to be possible using codes with conventional transversal gate constructions. } 

\section{Notation}

We denote axial rotations about the Pauli-$Z$ axis as
\begin{equation}
R(\theta) = \exp ( i \theta Z ) =  \cos(\theta) \id + i \sin(\theta)Z.
\end{equation}
If the angle is $\theta = \pi/2^{\ell}$ for integer $\ell$ then $R(\theta)$ belongs in the  $\ell^{\mathrm{th}}$ level of the Clifford hierarchy~\cite{CliffHier}.  Therefore, $R(\pi/8)$ is the $\pi/8$-phase gate, also known at the $T$ gate.  Unitaries inside the Clifford hierarchy are special because they can be realised using state-injection and a bounded number of appropriate magic states.  However, all the analysis in this paper holds for any $\theta$, even values  corresponding to unitaries and magic states not connected to the Clifford hierarchy.

We use $W(\theta)$ for the Hermitian operator $W(\theta) := R(\theta) X R(\theta)^\dagger = R(2 \theta)X$.  Note that $W(\pi/8)$ is a Clifford and plays a similar role to the Hadamard; it interchanges $X$ and $Y$ whereas the Hadamard interchanges $X$ and $Z$. The relevant magic states are eigenstates of $W(\theta)$ and sit on the equator of the Bloch sphere
\begin{equation}
\ket{R(\theta)} =  W(\theta) \ket{R(\theta)}  =  R(\theta) \ket{+}.
\end{equation}
More generally, when $U$ is a diagonal gate (acting on $n$ qubits) we use $\ket{U}:=U( \ket{+}^{\otimes n} )$.  In this notation, the familiar $T$ state is $\ket{R(\pi/8)} $.  We use CZ for control-Z and CCZ for control-control-Z. 

\section{Overview of new protocols}
\label{Overview}

\subsection{Protocols for $\pi/8$ phase gates}

The protocols presented here can be used both for distillation of $T$-states that can implement $\pi/8$ phase gates, or more generally for distillation of $\ket{R(\theta)}$ magic states for smaller angle rotations.   We begin by sketching the simple case of $T$-state distillation.  We say a protocol is an $n \rightarrow k$ protocol if it takes $n$ inputs and outputs $k$ magic states with some success probability (typically this probability approaches unity in the low noise limit).  Our two-step  protocols for $T$-state distillation are $3k+4 \rightarrow k$ protocols for even $k$ with quadratic reduction of noise.   The resource overhead of the protocol is roughly captured by $n/k = 3 + \frac{4}{k}$, which approaches 3 for large $k$. For $k=2$ we have a $10 \rightarrow 2$ protocol and so the protocol is very similar to the MEK proposal proposed by Meier, Eastin and Knill~\cite{Meier13}.  Therefore, compared to MEK, we reduce the $n/k$ overhead from 5 to 3 by going to larger blocks sizes (larger $k$).   Another class of protocols was proposed by Bravyi and Haah~\cite{Bravyi12}, which are  $3k+8 \rightarrow k$ protocols for even $k$ with quadratic reduction of noise.  The Bravyi-Haah protocols also have $n/k \rightarrow 3$ in the large protocol limit.  Both our protocols and the Bravyi-Haah protocols have the same asymptotic limit, but our protocols approach this limit faster.   For example, to achieve $n/k=4$ we can use modest size  $16 \rightarrow 4$ protocols, whereas the comparable Bravyi-Haah protocol is $32 \rightarrow 8$ with double the block size.  This effect becomes more pronounced as the protocol is concatenated.  

\begin{figure*}
	\begin{centering}
	\includegraphics{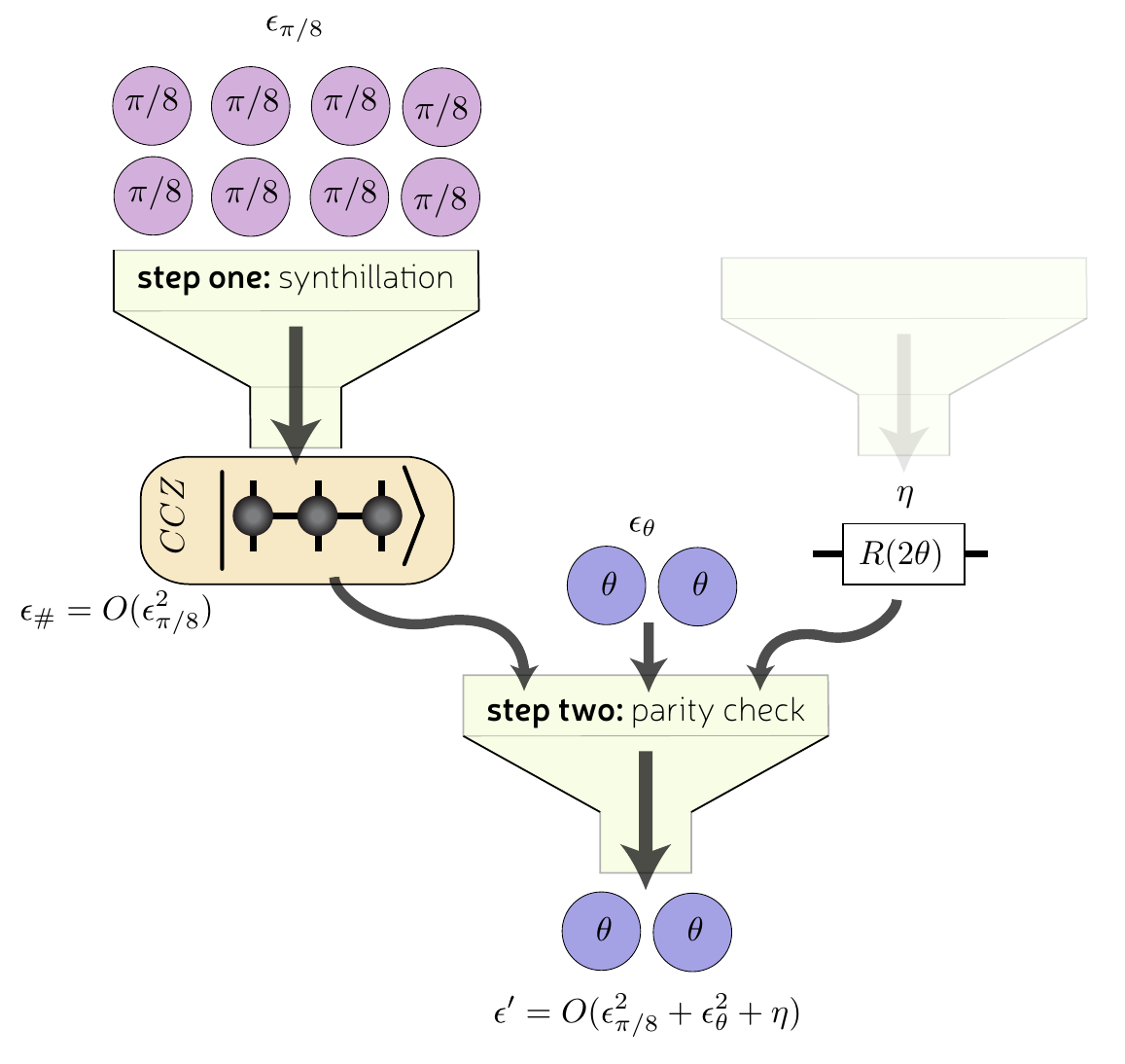}
	\caption{An illustration of how the two steps of our protocol are chained together for the special case $N=1$.  The combined process is described in Eq.~\eqref{ChainedEq} for general $N$. }
	\label{Fig:Chained}
		\end{centering}
\end{figure*}

Furthermore, our work presents a different approach to distillation as we break the process up into two steps, making use of a multi-qubit magic state resource.   Let us describe how this two-step process works, first considering the $10 \rightarrow 2$ case.  In the first step we prepare a ``pre-distilled'' magic state that can inject a CCZ (or Toffoli) gate.  It has been known for several years that a single CCZ magic state can be prepared from 8 $T$-states with quadratic error reduction~\cite{jones13b,eastin13}.  It is no longer appropriate to describe this process in simple $n \rightarrow k$ notation and so we introduce the more detailed description of these protocols as being
\begin{equation}
		\left\{ \begin{array}{c}
8 \\ 
\ket{R(\pi/8)} \\
\epsilon_{\pi/8} 
\end{array} \right\}
 \rightarrow \left\{ \begin{array}{c}
1 \\ 
\ket{CCZ} \\
O(\epsilon_{\pi/8}^2)
\end{array} \right\} .
\end{equation}
In this notation, the left hand side gives the input resources and the right hand size gives the output resources.  The top line gives the quantity of resources, the middle describes the species of magic state and the bottom line gives the infidelity.  Our second step, which was not previously known, is to notice that a single $\ket{CCZ}$ resource can be used to check the parity in a pair of $\ket{R(\pi/8)} $ states, which implements the transform
\begin{equation}
\left\{ \begin{array}{cc} 2 &  1 \\ 
\ket{R(\pi/8)} & \ket{CCZ}\\
\epsilon_{\pi/8} & \epsilon_{\#}
\end{array} \right\}
\rightarrow \left\{ \begin{array}{c}
2 \\ 
\ket{R(\pi/8)} \\
O(\epsilon_{\pi/8}^2 + \epsilon_{\#})
\end{array} \right\} .
\end{equation}
Chaining these steps together so that $\epsilon_{\#}=O(\epsilon_{\pi/8}^2)$ yields
\begin{equation}
\left\{ \begin{array}{c}
10 \\ 
\ket{R(\pi/8)} \\
\epsilon_{\pi/8} 
\end{array} \right\}
\rightarrow 
\left\{ \begin{array}{cc}
2  &  1 \\
\ket{R(\pi/8)} & \ket{CCZ} \\
\epsilon_{\pi/8}  & O(\epsilon_{\pi/8}^2)
\end{array} \right\}
\rightarrow \left\{ \begin{array}{c}
2 \\ 
\ket{R(\pi/8)} \\
O(\epsilon_{\pi/8}^2)
\end{array} \right\} ,
\end{equation}
or simply $10 \rightarrow 2$ for short.   

Next we outline the two-step process for larger block protocols.   Instead of using $\ket{CCZ}$ resources, larger block protocols will use a resource that we denote as $\ket{CCZ_{\#N}}$.  This resource can inject a gate $CCZ_{\#N}$, which is $N$ copies of the CCZ gate all sharing one control qubit in common, and the relevant magic state is simply
\begin{equation}
	\ket{CCZ_{\#N}}:=CCZ_{\#N} (\ket{+}^{\otimes 2N+1}).
\end{equation}
In the first step, we borrow results on synthillation~\cite{campbell17,campbell17b} that provide protocols implementing
\begin{equation}
\left\{ \begin{array}{c}
4N+4 \\ 
\ket{R(\pi/8)} \\
\epsilon_{\pi/8} 
\end{array} \right\}
\rightarrow \left\{ \begin{array}{c}
1 \\ 
\ket{CCZ_{\#N}} \\
O(\epsilon_{\pi/8}^2)
\end{array} \right\} .
\end{equation}
This protocol is described in the section "$U_{N\#}$ family" of Ref.~\cite{campbell17b}.  To avoid repetition, here we simply treat this synthillation routine as a black box with known properties.  Rather, here we focus on the second step --- the main technical contribution of this work --- by showing that a pre-distilled $\ket{CCZ_{\#N}}$ can be used to parity check on $2N$ magic states
\begin{equation}
\left\{ \begin{array}{cc} 2N &  1 \\ 
\ket{R(\pi/8)} & \ket{CCZ_{\#N}    }\\
\epsilon_{\pi/8} & \epsilon_{\#}
\end{array} \right\}
\rightarrow \left\{ \begin{array}{c}
2N \\ 
\ket{R(\pi/8)} \\
O(\epsilon_{\pi/8}^2 + \epsilon_{\#})
\end{array} \right\} .
\end{equation}
Chaining steps one and two, so that $\epsilon_{\#}=O(\epsilon_{\pi/8}^2)$, we obtain a family of protocols for all integer $N$
\begin{equation}
	\label{ChainedEq}
\left\{ \begin{array}{c} 6N+4  \\ 
\ket{R(\pi/8)} \\
\epsilon_{\pi/8} 
\end{array} \right\}
\rightarrow \left\{ \begin{array}{cc}
2N &  1\\ 
\ket{R(\pi/8)} & \vert CCZ_{\#N} \rangle \\
\epsilon_{\pi/8} & O(\epsilon_{\pi/8}^2)
\end{array} \right\}
\rightarrow \left\{ \begin{array}{c}
2N \\ 
\ket{R(\pi/8)} \\
O(\epsilon_{\pi/8}^2)
\end{array} \right\} ,
\end{equation}
or simply $6N+4 \rightarrow 2N$ for short.  Alternatively, using $k=2N$ we have a family of $3k+4 \rightarrow k$ protocols with even $k$.  So we have returned to the choice of symbols used by Bravyi and Haah who found $3k+8 \rightarrow k$ protocols with even $k$.  Fig.~\ref{Fig:Chained} illustrates this for the $N=1$ ($k=2$) case. This completes the outline of our protocols for $T$-state distillation. 

\subsection{Protocols for general phase gates}

Next, we describe a family of protocols for other equatorial magic states $\ket{R(\theta)}$.  Step one will remain the same, again making use of synthillation of $\ket{CCZ_{\#N}}$ resource states.  Step two generalises to 
\begin{align}& \left\{ \begin{array}{ccc} 2N &  1 & N  \\ 
\ket{R(\theta)} & \ket{CCZ_{\# N}} & R(2 \theta) \\
\epsilon_{\theta} & \epsilon_{\#} & \eta
\end{array} \right\} \rightarrow   \left\{ \begin{array}{c}
2N \\ 
\ket{R(\theta)} \\
O(\epsilon_{\theta}^2+\epsilon_{\#} + \eta)
\end{array} \right\} .
\end{align}
When $R(2 \theta)$ appears uncluttered by a ket, as it does above, it refers to inputting a phase gate $R(2 \theta)$ rather than a magic state.  Most interesting is the case when $R(2 \theta)$ sits in the Clifford hierarchy and so can be injected using appropriate magic states.  Proving the validity of the above mapping is at the core of this paper.  Chaining this with synthillation yields an overall protocol
\begin{align}
 \left\{ \begin{array}{@{\hspace{0pt}}ccc@{\hspace{0pt}}} 2N &  4N+4 & N  \\ 
\ket{R(\theta)} & \ket{R(\pi/8)} & R(2 \theta) \\
\epsilon_{\theta} & \epsilon_{\pi/8} & \eta
\end{array} \right\} &\rightarrow \left\{ \begin{array}{@{\hspace{0pt}}ccc@{\hspace{0pt}}}
2N &  1 & N \\ 
\ket{R(\theta)} & \vert CCZ_{\# N} \rangle & R(2 \theta) \\
\epsilon_{\theta} & O(\epsilon_{\pi/8}^2) &  \eta
\end{array} \right\}
 \rightarrow \left\{ \begin{array}{@{\hspace{0pt}}c@{\hspace{0pt}}}
2N \\ 
\ket{R(\theta)} \\
O(\epsilon_{\theta}^2 + \epsilon_{\pi/8}^2 + \eta)
\end{array} \right\} .
\end{align}
It is important to notice that there is no noise reduction in $\eta$.  Therefore, it is crucially important they are predistilled (e.g. $\eta \sim \epsilon^2$) and so we refer to the rotations $R(2 \theta)$ as pivotal. Despite the need for high fidelity pivotal rotations, other protocols have used pivotal rotations and found significant reductions of resource costs compared against using gate-synthesis.   For instance, the $N=1$ two-step protocol has an identical resource cost to the protocols introduced by Campbell and O'Gorman~\cite{smallAngle}.   Pivotal rotations (though not under this name) played a similar role in the protocols proposed by Duclos-Cianci and Poulin~\cite{duclos15}, which in our notation can be described as
\begin{align}
 \left\{ \begin{array}{ccc} 2N &  8N+4 & N \\ 
\ket{R(\theta)} & \ket{R(\pi/8)} & R(2 \theta) \\
\epsilon_{\theta} & \epsilon_{\pi/8} & \eta
\end{array} \right\} 
 \rightarrow \left\{ \begin{array}{c}
2N \\ 
\ket{R(\theta)} \\
O(\epsilon_{\theta}^2 + \epsilon_{\pi/8}^2 + \eta)
\end{array} \right\} .
\end{align}
Note that for the majority of their paper, Duclos-Cianci and Poulin only discuss the $N=1$ case, but they do sketch the higher $N$ case later in the paper.   

One could say our protocols are compressed as they essentially give a slight compression in $T$-cost of the Duclos-Cianci and Poulin protocols~\cite{duclos15}.  Our protocols also have very different inner workings.  This provides a new perspective on magic state distillation, but also the two-step feature has a potentially significant practical advantage.   The exotic resources are higher value since the $R(2 \theta)$ pivotal rotation and $\ket{R(\theta)}$ resources are more difficult to prepare than standard $\ket{R(\pi/8)} $ magic states.  However, in the two-step protocols, one does not risk using the exotic resources until the first step has succeeded.  This contrasts, with both the Duclos-Cianci-Poulin protocols and Campbell-O'Gorman protocols for which all the resources are committed at the same time, with a single error anywhere leading to loss of all resources.

\begin{figure*}
	\centering
	\includegraphics[width=\columnwidth]{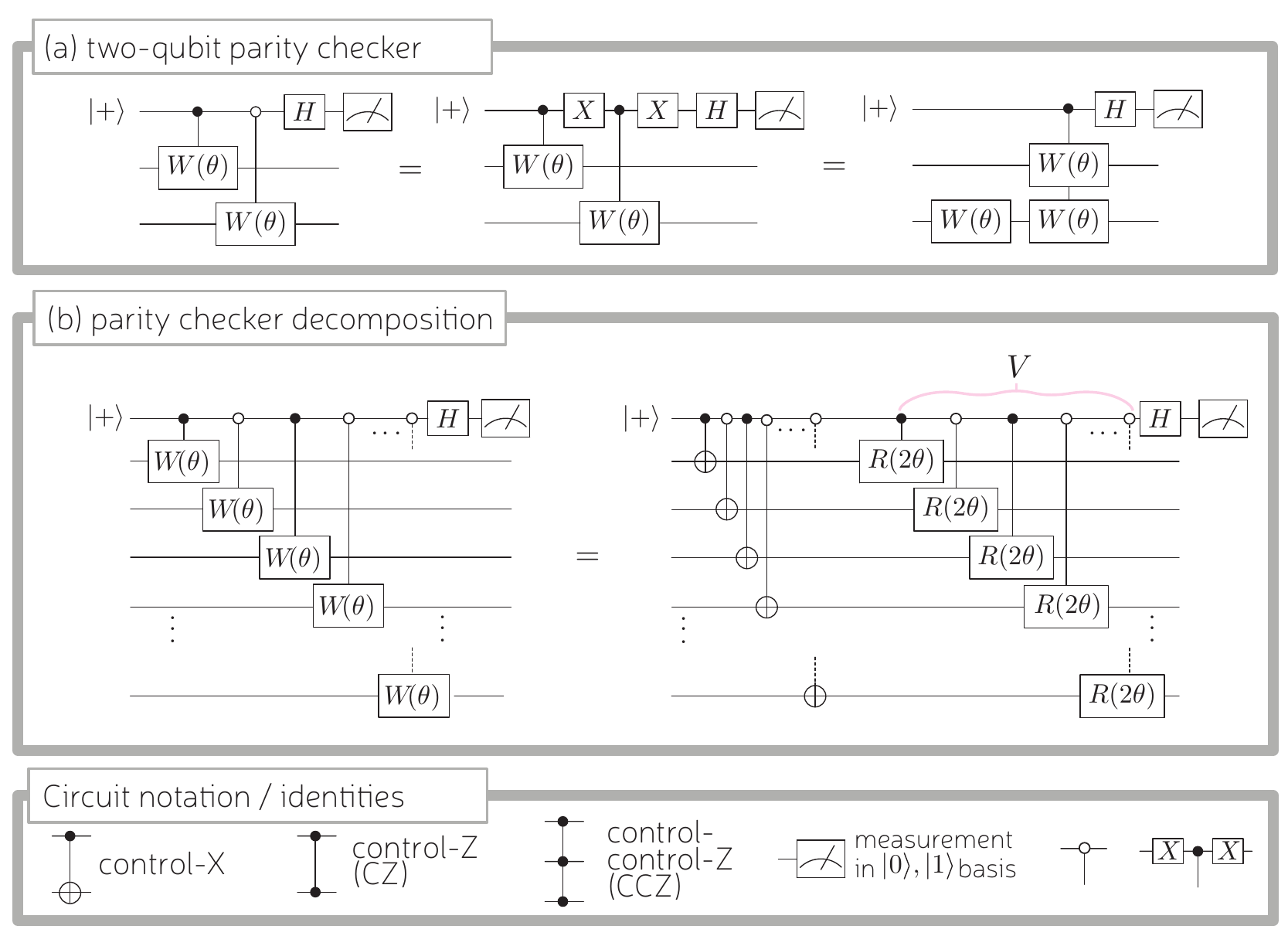}
	\caption{Gadgets for measuring parity in the $W(\theta)$ basis.  In (a) we illustrate how control-$W(\theta)$ gates are used to measure parity. In (b) we show how the control-$W(\theta)$ gates can be decomposed into CNOT and control-phase gates. In the text we describe this phase-gate circuit by $V$ (see Eq.~\eqref{eqn_V}), which can be broken up into a product of $U_j$ gates (see Eq.~\eqref{eqn_Uj}).}
	\label{Fig:parity_check}
\end{figure*}

\section{Implementing step two}
\label{Sec_two_step}

Here we show how to use $CCZ$ circuits, implemented using synthillation, to perform a parity check. We will construct circuits that measure the parity of $2N$ qubits in the basis of $W(\theta)$.  Using an ancilla and a control-$W(\theta)^{\otimes 2N}$ gate would achieve this, but a smaller resource overhead is needed if we instead use the parity check gadgets in Fig.~\ref{Fig:parity_check}. We use a combination of control-$W(\theta)$ that triggers when the control bit is in the $\ket{1}$ state with an unconventional control-$W(\theta)$ (shown with open circles) that triggers when the control bit is in the $\ket{0}$ state.  This parity check gadget measures in the $W(\theta)$ basis, but with an additional $W(\theta)$ rotation  on half the qubits.  This resulting Kraus operator for the desired even parity outcome is
\begin{align}
\label{ProjectionKraus} \nonumber
    K & =\frac{1}{2} (\id^{\otimes N} \otimes W(\theta)^{\otimes N}  +   W(\theta)^{\otimes N} \otimes \id^{\otimes N} ) \\ 
    & =\frac{1}{2} (\id^{\otimes N}  \otimes W(\theta)^{\otimes N} ) (\id^{\otimes 2N}   +   W(\theta)^{\otimes 2N}  ) .
\end{align}
We assume the noisy $\ket{R(\theta)}$ magic states are diagonal in the $W(\theta)$ basis, and so the additional  $W(\theta)$ rotations have no effect.  The assumption of diagonal noise is mild and can be removed at the expense of hideously complicating the noise analysis (see e.g. App D of Ref.~\cite{smallAngle}).

 Using $W(\theta)=R(2\theta)X$ it follows that this parity measurement circuit can be split into a sequence of control-$X$ gates followed by a phase gate circuit (see Fig.~\ref{Fig:parity_check}b).  Algebraically, this  phase gate circuit is
\begin{align}
\label{eqn_V}
	V & = \prod_{j=1,\ldots,N} \left[ \kb{0}{0}_0  R_{2j}(2\theta) + \kb{1}{1}_0  R_{2j-1}(2\theta) \right] ,
\end{align} 
where the subscripts denote which qubits the rotations act on, with qubit labels running from $0$ to $2N$.  Using the shorthand
\begin{align}
   U_j & := \kb{0}{0}_0 R_{2j}(2\theta) + \kb{1}{1}_0  R_{2j-1}(2\theta),
\end{align} 
we have 
\begin{align}
V & = \prod_{j=1,\ldots, N} U_j .
\end{align} 
To recap, given $\ket{R(\theta)}$ of error rate $\epsilon$ and the ability to implement $V$  we can parity check in the $W(\theta)$ basis, outputting $\ket{R(\theta)}$ of error rate $O(\epsilon^2)$ with some probability $p=1-O(\epsilon)$.

Next, we show how to implement $V$ with some pre-distilled resources.  First, we use the decomposition $R(2\theta)=\cos(2 \theta) \id + i \sin(2 \theta)Z$ to expand out $U_j$ as
\begin{align}
\label{eqn_Uj}
	U_j  = &  \kb{0}{0}_0 (\cos(2 \theta) \id + i \sin(2 \theta)Z_{2j} )+ \kb{1}{1}_0 (\cos(2 \theta) \id + i \sin(2 \theta)Z_{2j-1} ).
\end{align} 
Collecting the $\cos$ and $\sin$ terms, we have
\begin{align}
U_j  = &  \cos(2 \theta) \id + i \sin(2 \theta) [ \kb{0}{0}_0 Z_{2j} + \kb{1}{1}_0 Z_{2j-1} ]  \\ \nonumber 
  = &  \cos(2 \theta) \id + i \sin(2 \theta) M_j ,
\end{align}
where we have introduced further shorthand
 \begin{align} \nonumber
   M_j & :=  \kb{0}{0}_0 Z_{2j} + \kb{1}{1}_0 Z_{2j-1} \\ \nonumber
   & = Z_{2j} ( \kb{0}{0}_0+ \kb{1}{1}_0 Z_{2j}  Z_{2j-1}  )  \\
  & = Z_{2j}  CZ_{0,2j} CZ_{0,2j-1} ,
 \end{align}	
which is unitary, Hermitian and Clifford.  
\begin{figure*}
	\centering
	\includegraphics[width=\columnwidth]{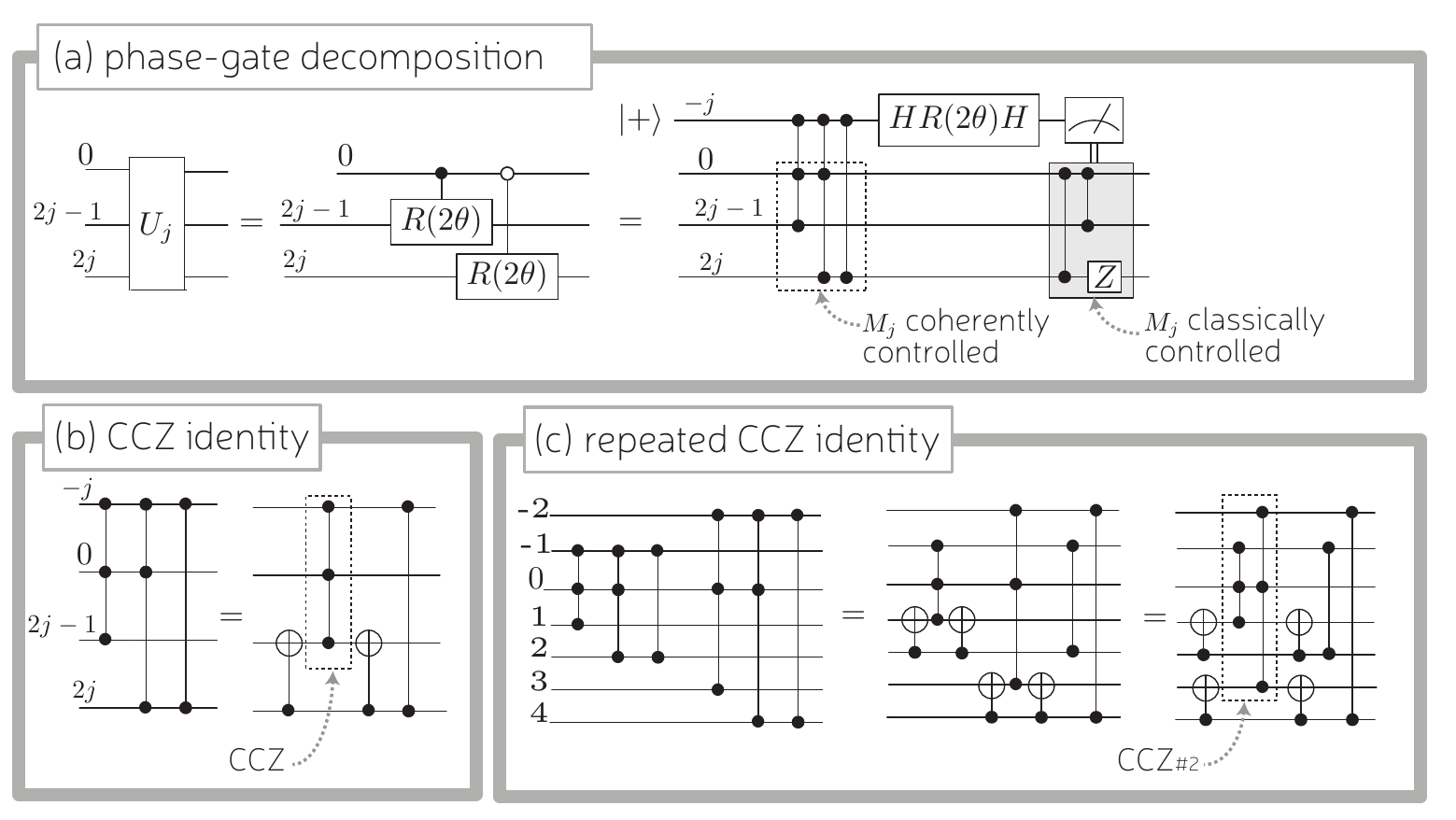}
	\caption{In (a) we show how $U_j$ (a pair of control phase gates) can be implemented using an ancilla, a control-$M_j$ gate and a pivotal rotation.  An algebraic proof of this equivalence starts at Eq.~\eqref{eqn_Uj} and ends after Eq.~\eqref{end_Uj_proof}.  In (b) we show how a pair of CCZ gates, which share a pair of controls, is Clifford equivalent to only a single CCZ gate.  In (c) we uses this identity twice to simplify a more complex circuit down to a $CCZ_{\# 2}$.  In general, one finds that a sequence of $N$ control-$M_j$ can be simplified to a $CCZ_{\# N}$ circuit (see Eq.~\eqref{tildeV} and Eq.~\eqref{tildeV2}).}
	\label{phase_gate_decompositions}
\end{figure*}

Next we show that each $U_j$ can be implemented with access to a single $\ket{+}$ ancilla, a control-$M_j$ unitary and  a  $R(2\theta)$ rotation (see Fig.~\ref{phase_gate_decompositions}a).   We prepare the ancilla in the $\ket{+}$ state and use it as the control qubit for the control-$M_j$ unitary, which gives the state $\ket{0}\ket{\psi}  + \ket{1}(M_j\ket{\psi})  $.  Next we rotate the ancilla by $HR(2\theta)H$ and measure in the computational basis.  This is equivalent to measuring with projections
\begin{align}  
  \bra{0}   HR(2\theta) H  & =  \cos(2\theta)\bra{0} + i \sin(2\theta) \bra{1} \\
  \bra{1}  HR(2\theta)H & =  \cos(2\theta)\bra{1} + i \sin(2\theta) \bra{0} .
\end{align}
In the eventuality of a ``+1'' outcome, we find
\begin{align}
 \ket{0}\ket{\psi}  + \ket{1}(M_j\ket{\psi})   \rightarrow (\cos(2\theta) \id + i \sin(2 \theta)M_j ) 	\ket{\psi}  = U_j	\ket{\psi} ,
\end{align} 
as desired.  However, when a ``1'' outcome is measured we have
 \begin{align}
 \label{end_Uj_proof}
  \ket{0}\ket{\psi}  + \ket{1}(M_j\ket{\psi})   \rightarrow (i \sin(2 \theta) \id + \cos(2\theta) M_j ) 	\ket{\psi}  = M_j U_j	\ket{\psi} .
 \end{align} 
We see that an $M_j$ gate will correct for the different measurements outcomes, and since $M_j$ is Clifford this does not contribute to the resource cost.  This completes the proof of the identity in Fig.~\ref{phase_gate_decompositions}a.

Similarly we chain together $N$ such circuits, assuming we have access to $N$ copies of $R(2\theta)$ and the circuit
\begin{align}
\label{tildeV}
\tilde{V} & = \prod_{j=1,\ldots, N}  ( \kb{0}{0}_{-j} +  \kb{1}{1}_{-j} M_j) ,\\
& = \prod_{j=1,\ldots, N}  CZ_{-j,2j} CCZ_{-j,0,2j} CCZ_{-j,0,2j-1}  .
\end{align}	
This is composed of a sequence of control-$M_j$ gates, where each gate is controlled from a different ancillary qubit. We have labelled these new ancilla with negative integers from $-1$ to $-N$.  Each control-$M_j$ consists of a CZ gate and two CCZ gates.  On first inspection this seems to imply $2N$ CCZ gates are needed.  However, using the identity in Fig.~\ref{phase_gate_decompositions}b we see a pair of CZZ gates can sometimes be realised using a single CCZ.   Note this identity only works because the pair of CCZ gates share two control bits in common.  Applying this identity repeatedly (see e.g. Fig.~\ref{phase_gate_decompositions}c) can reduce the circuit to one using only $N$ CCZ gates.  Algebraically, the identity is
\begin{align}
\label{tildeV2}
\tilde{V} & =\left( \prod_{j=1}^N  CX_{2j,2j-1}  \right)   \left( \prod_{j=1}^N  CCZ_{-j,0,2j-1}  \right) \left( \prod_{j=1}^N  CX_{2j,2j-1}  \right)  \left(  \prod_{j=1}^N  CZ_{-j,2j}  \right)   ,
\end{align}	
where $CX_{c,t}$ is a control-$X$ gate with control qubit $c$ and target qubit $t$.   The resource intensive part is the non-Clifford component of $N$ CCZ gates all sharing one single control qubit in common (qubit $0$).  Here we denote such a circuit as CCZ$_{\# N}$, which has been elsewhere called Tof$_{\# N}$.  For these gates, the problem of optimal synthesis into CNOT + $T$ gates has been solved and the circuit requires $4N+3$ $T$ gates (see Example IV.2. of Ref.~\cite{campbell17b}).  Recall that we are requiring that CCZ$_{\# N}$ is predistilled to a higher fidelity.  The most efficient known method to achieve this is to use the synthillation protocol that can perpare CCZ$_{\# N}$ using only $(4N+4)$ $T$-states of $\epsilon_{\pi/8}$ error rate, and so this is the first-step of our two-step protocol. 

\begin{figure*}
	\includegraphics[width=\columnwidth]{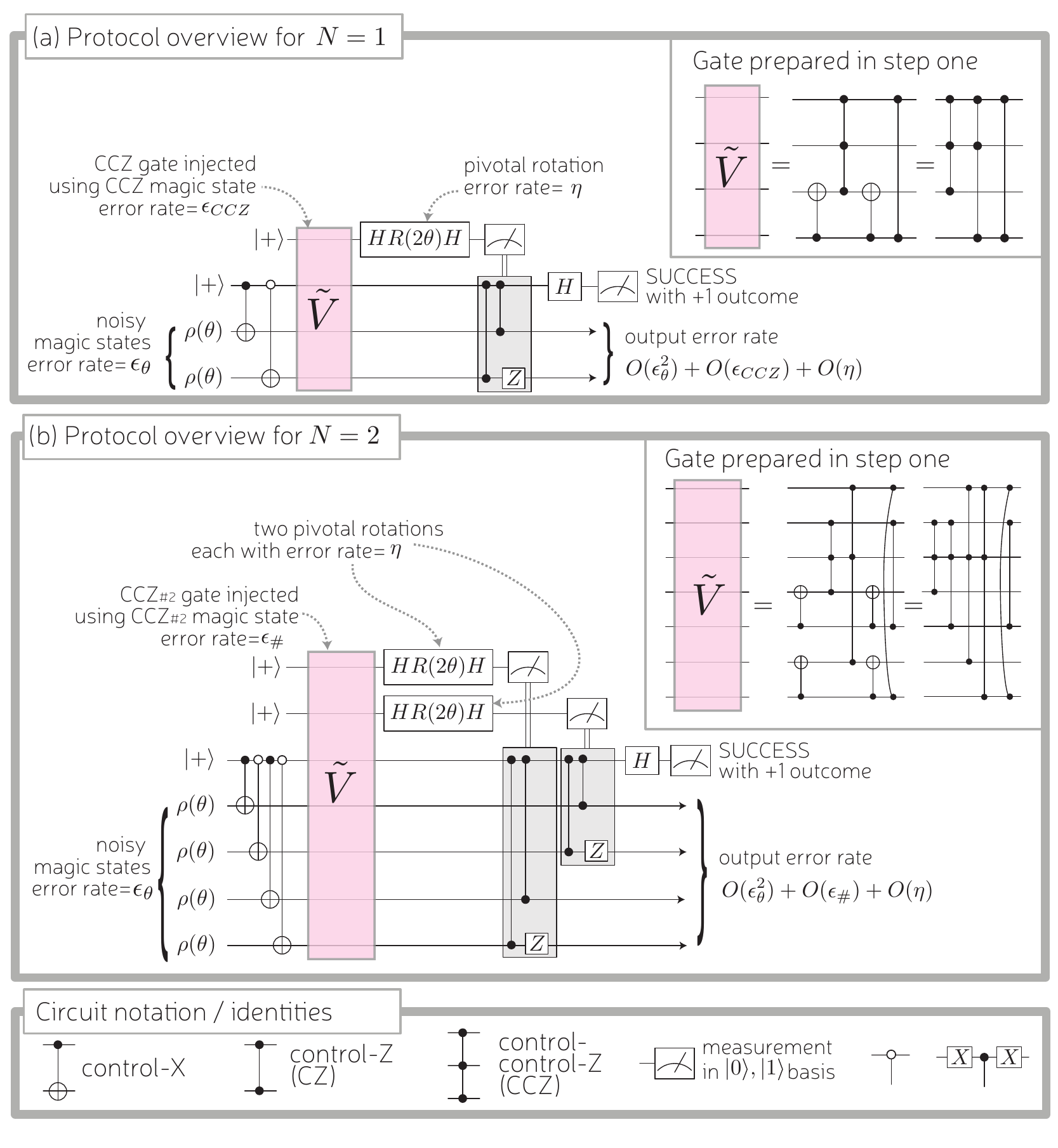}
	\caption{An overview of the second-step of our protocol for $N=1$ and $N=2$, which can be extended to any integer $N$.  The first-step is to use synthillation to inject the non-Clifford gates shown in the pink section of the circuit.  The only other non-Clifford elements of the circuits are the pivotal rotations $R(\theta)$.  Input to the circuit are $2N$ mixed states $\rho(\theta)$, which are $\ket{R(\theta)}$ magic states with $\epsilon$ phase noise.  In the absence of noise, the circuit measures the parity in the $W(\theta)$ bases.  Therefore, when we see a SUCCESS event, the $\rho(\theta)$ states are output with quadratically reduced noise.  In the event of a FAILURE, we discard the qubits and attempt again.} 
	\label{Fig:megaMEK}
\end{figure*}

We have demonstrated how step two works using a series of circuit identities (for any integer $N$).  For completeness, we show in Fig.~\ref{Fig:megaMEK} how these circuit identities plug together for $N=1$ and $N=2$.  The circuits could be further expanded by replacing the non-Clifford gate $CCZ_{\#N}$ with the magic state $\ket{CCZ_{\#N}}$ and the appropriate Clifford injection circuit.

\section{Noise analysis}
\label{Sec::noise}

This section presents a performance analysis for one round of our protocols. Subsection~\ref{Sec::Numerical_noise} reports some results of numerical simulations for smaller size protocols with $N=1$ and $N=2$.  Subsection~\ref{Sec::Analytic_noise} focuses on providing a simple, yet rigorous derivation, of analytic upper bounds on output noise.  The analytic results hold for all $N$, but are loose and actual performance will be much better than analytically bounded. 

\subsection{Numerical analysis}
\label{Sec::Numerical_noise}

We performed full state vector numerical simulations using IBM's QISKit (code available as ancillary file).  We simulated the effect of leading order errors for circuits with $N=1$ and $N=2$.  The output error probabilities are for a single qubit with other output qubits traced out.  Numerical results were independent of which output qubit is chosen and independent of $\theta$ (up-to numerical accuracy of 2 significant figures). 

For $N=1$, we found that 
\begin{align}
\label{Numerical_error_out}
\epsilon' & =  8 \epsilon_{\pi/8}^2 +   \epsilon_{\theta}^2 + \frac{1}{4} \eta + O(\epsilon_{\pi/8}^3,\epsilon_{\theta}^2,\eta^2,\ldots ).
\end{align}
The leading order coefficients for output error are identical to those for the MEK$_{k}$ protocols proposed by Campbell and O'Gorman (themselves a modified form of MEK) and so it seems that our new protocols (with $N=1$) perform identically in this regard.   We have a slight performance advantage in terms of success probability due to the two step nature.   We found
\begin{align}
p_{\mathrm{synth}} & = 1 - 8  \epsilon_{\pi/8}  + O(\epsilon_{\pi/8}^2), \\ \nonumber
p_{\mathrm{parity}} & = 1 - 2 \epsilon_{\theta} - \frac{1}{2} \eta + O(\epsilon_{\pi/8}^2, \epsilon_{\theta}^2, \eta^2,  \ldots ),
\end{align}  
where $p_{\mathrm{synth}} $ is the probability of step one succeeding and $p_{\mathrm{parity}}$ is the probability of step two succeeding.  To leading order, the previous MEK$_{k}$ protocols had a success probability equal to $p_{\mathrm{mek}} = p_{\mathrm{synth}}p_{\mathrm{parity}}$.  Here, we don't commit to the second step until the first step is successful, which will lead to superior rates of generating magic states. For the setting $\theta = \pi/8$, the protocol simplifies to a $10 \rightarrow 2$ protocol with $\epsilon' = 9 \epsilon_{\pi/8}^2  + O(\epsilon_{\pi/8}^3 )$ and overhead $n/k=10/2=5$, very similar to the original MEK protocol. 

For $N=2$, we found that 
\begin{align}
\label{Numerical_error_out2}
\epsilon' & =  16  \epsilon_{\pi/8}^2 +  3 \epsilon_{\theta}^2 + \frac{1}{4} \eta + O(\epsilon_{\pi/8}^3,\epsilon_{\theta}^2,\eta^2,\ldots ), \\ 
p_{\mathrm{synth}} & = 1 - 12  \epsilon_{\pi/8}  + O(\epsilon_{\pi/8}^2), \\
p_{\mathrm{parity}} & = 1 - 6 \epsilon_{\theta} -  \eta + O(\epsilon_{\pi/8}^2, \epsilon_{\theta}^2, \eta^2, \ldots),
\end{align}
By going to $N=2$, we incur only mildly worse constant prefactors, but gain a significant efficiency improvement in terms of magic states output per input.   For the setting $\theta = \pi/8$, the protocol simplifies to a $16 \rightarrow 4$ protocol with $\epsilon' = 19 \epsilon_{\pi/8}^2  + O(\epsilon_{\pi/8}^3 )$ and overhead $n/k=16/4=4$.  To obtain the same $n/k$ overhead using Bravyi-Haah protocols (which are limited to $\theta = \pi/8$), we need to go to a larger size $32 \rightarrow 8$ protocol with $\epsilon' = 25 \epsilon_{\pi/8}^2  + O(\epsilon_{\pi/8}^3 )$.  This confirms that our protocols can obtain similar resource overheads with a smaller scale quantum computer, and without any sacrifice in terms of error suppression or success probability.

\subsection{Algebraic analysis}
\label{Sec::Analytic_noise}

Here we take an analytic approach.  We do not know of an analytic method of determining the exact expressions for $\epsilon'$, but can prove a rigourous upper bound using standard norm inequalities. Actual performance will be much better than proven here.  We begin by considering the effect of noise on the input states
\begin{equation}
	 \rho(\theta) := (1- \epsilon_{\theta} ) \kb{R(\theta)}{R(\theta)}+ \epsilon_{\theta}  Z\kb{R(\theta)}{R(\theta)}Z.
\end{equation}
We extend later to account for CCZ$_{\# N}$ noise and pivotal rotation noise, but when these components work perfectly the circuit implements
	 \begin{equation}
	  \rho(\theta)^{\otimes 2N} \rightarrow  \mathcal{E}( \rho(\theta)^{\otimes 2N} ) = K \rho(\theta)^{\otimes 2N} K^\dagger
	\end{equation}
where $K$ is the parity projecting Kraus operator introduced in Eq.~\eqref{ProjectionKraus}.  Rather than the whole multi-qubit output, we are interested in the fidelity of a single output qubit, and so introduce the channel
	 \begin{equation}
     \mathcal{E}_i ( \rho(\theta)^{\otimes 2N} ) = \mathrm{tr}_i  [ K \rho(\theta)^{\otimes 2N} K^\dagger] ,
\end{equation}
where $\mathrm{tr}_i [ \cdots ]$ is the partial trace over all but the $i^{\mathrm{th}}$ qubit.  The output of this channel is the unnormalised state
	 \begin{equation}
	 \label{ChanOutEq}
\mathcal{E}_i ( \rho(\theta)^{\otimes 2N} ) =   p_{\mathrm{good}} \kb{R(\theta)}{R(\theta)} + p_{\mathrm{bad}}  Z \kb{R(\theta)}{R(\theta)}Z .
\end{equation}
The renormalisation constant $p_{\mathrm{good}}+p_{\mathrm{bad}}$ is the probability of the parity check yielding a ``+1" outcome. When the parity check process is error-free, this occurs whenever the input states contains no errors or an even number of errors, and so
\begin{align}
	p_{\mathrm{good}}+p_{\mathrm{bad}} & =  \frac{1}{2}(1 + (1-2\epsilon_{\theta})^{2N})
\end{align}
The term $p_{\mathrm{bad}}$ is the probability of an error on $i^{\mathrm{th}}$ qubit and an odd number of errors on the remaining $2N-1$ qubits
\begin{align}
\label{Pbad_inequality}
	p_{\mathrm{bad}} & = \epsilon_{\theta}  \sum_{j=1}^{N}  \genfrac(){0pt}{1}{2N-1}{2j - 1}    \epsilon_{\theta}^{2j-1} (1- \epsilon_{\theta})^{2N-2j}  , \\ \nonumber
	& = \frac{\epsilon_{\theta}}{2}(1 - (1-2\epsilon_{\theta})^{2N-1}) , \\ \nonumber 
	& < (2N-1) \epsilon_{\theta}^2 ,
\end{align}
where in the first line we have a binomial coefficient.  The inequality follows from Bernoulli's inequality.  This shows quadratic noise suppression in $\epsilon_{\theta}$.

Next, we account for $\epsilon_{\#}$ noise in the CCZ$_{\#N}$ gate.  We can write the corresponding magic state as
\begin{equation}
		\rho_{\#N} = (1 - \epsilon_{\#}) \Psi_{\#N} + \epsilon_{\#} \sigma_{\#N},
\end{equation}
where 
\begin{equation}
  \Psi_{\#N} = \kb{CCZ_{\#N}}{CCZ_{\#N}},
\end{equation}
and $\sigma_{\#N}$ carries some $Z$ noise.  We define $\mathcal{F}$ as the channel describing the action of the whole  circuit (including implicit injection gadget for $CCZ_{\#N}$), assuming ideal pivotal rotations, acting on $\rho_{\#N}$ and  $\rho(\theta)^{\otimes 2N}$.  We will use that $\Psi_{\#N}$ leads to a parity 
\begin{equation}
     \mathcal{F} ( \Psi_{\#N} \otimes \rho(\theta)^{\otimes 2N}  ) = \mathcal{E}( \rho(\theta)^{\otimes 2N}  )
\end{equation}
and by linearity of $\mathcal{F}$ we deduce
\begin{align}
& \mathcal{F} ( \rho_{\#N} \otimes \rho(\theta)^{\otimes 2N}  ) =  (1 - \epsilon_{\#})  \mathcal{E}( \rho(\theta)^{\otimes 2N}  )   + \epsilon_{\#}  \mathcal{F} ( \sigma_{\#N} \otimes \rho(\theta)^{\otimes 2N}  ) .
\end{align}
Again, we are interested in only the single output qubit, and so introduce $\mathcal{F}_i = \mathrm{tr}_i \mathcal{E}_i $, which straightforwardly yields
\begin{equation}
\mathcal{F}_i ( \rho_{\#N} \otimes \rho(\theta)^{\otimes 2N}  ) = (1 - \epsilon_{\#})  \mathcal{E}_i( \rho(\theta)^{\otimes 2N}  ) + \epsilon_{\#}  \mathcal{F}_i ( \sigma_{\#N} \otimes \rho(\theta)^{\otimes 2N}  ) .
\end{equation}
This yields a single qubit state of the form in Eq.~\eqref{ChanOutEq} with new parameters $p'_{\mathrm{good}}$ and $p'_{\mathrm{bad}}$, which are tricky to exactly calculate but can again be bounded.  The joint probability $p'_{\mathrm{good}} + p'_{\mathrm{bad}}$ can be lower bounded by assuming $\mathcal{F}_i ( \sigma_{\#N} \otimes \rho(\theta)^{\otimes 2N}  )=0$ and so
\begin{align}
	   p'_{\mathrm{good}} + p'_{\mathrm{bad}} & \geq (1-\epsilon_{\#})(p_{\mathrm{good}} + p_{\mathrm{bad}})
	   \end{align} 
The  $p'_{\mathrm{bad}}$ term can be upper bounded by considering the worst-case scenario that $\mathcal{F}_i ( \sigma_{\#N} \otimes \rho(\theta)^{\otimes 2N}  )$ leads to a logical error with unit probability, and so
  \begin{align}
	   p'_{\mathrm{bad}} & \leq (1-\epsilon_{\#}) p_{\mathrm{bad}} + \epsilon_{\#} <(2N-1) (1-\epsilon_{\#})\epsilon_{\theta}^2 + \epsilon_{\#} ,
\end{align}
where the second inequality follows from Eq.~\eqref{Pbad_inequality}.  These bounds are very loose and overestimate $p'_{\mathrm{bad}}$ by quite a lot.  Nevertheless, they are simple to obtain and rigorous.   

Next, we further consider phase noise on the pivotal rotation, each failing with probability $\eta$.  In other words, all pivotal rotations act perfectly with probability $(1-\eta)^N$.  Therefore, the channel implemented is not $\mathcal{F}_i$ but something of the form
\begin{equation}
		\mathcal{G}_i = (1-\eta)^N \mathcal{F}_i +(1-(1- \eta)^N) \mathcal{F}'_i ,
\end{equation}
where $\mathcal{F}'_i$ is the noisy part of the channel with diamond norm not exceeding unity.  Therefore,
\begin{align}
	\mathcal{G}_i (  \rho_{\#N} \otimes \rho(\theta)^{\otimes 2N}  )   = &  (1-\eta)^N( p'_{\mathrm{good}}  \kb{R(\theta)}{R(\theta)} +  p'_{\mathrm{bad}} Z\kb{R(\theta)}{R(\theta)}Z )   \\ \nonumber
	& + (1-(1-\eta)^N ) \mathcal{F}'_i (  \rho_{\#N} \otimes \rho(\theta)^{\otimes 2N}  ).
\end{align}
The worst case scenario is that $\mathcal{F}'_i $ always generates an error,  adding a $(1-(1-\eta)^N )$ contribution to the error term.  Therefore, after renormalising the error probability is bounded by
\begin{align} \nonumber
   \epsilon' & \leq \frac{(1-\eta)^N p'_{\mathrm{bad}}  + (1-(1- \eta)^N)}{(1-\eta)^N(p_{\mathrm{good}}' + p_{\mathrm{bad}}') + (1-(1- \eta)^N) } \\ \nonumber
& \leq \frac{(1-\eta)^N((2N-1) (1-\epsilon_{\#})\epsilon_{\theta}^2 + \epsilon_{\#} ) + (1-(1- \eta)^N) }{(1-\eta)^N(1-\epsilon_{\#})(p_{\mathrm{good}} + p_{\mathrm{bad}}) + (1-(1- \eta)^N) }
\end{align}
The result scales as $O(\epsilon_{\#})$, but this error rate is itself the output of performing the synthillation protocol using noisy $\ket{R(\pi/8)} $-states of error rate $\epsilon_{\pi/8}$.  In particular, Eq.~(128) of Ref.~\cite{campbell17b} shows that
\begin{equation}
\epsilon_{\#} \leq 1 - \frac{2(1-\epsilon_{\pi/8})^{4N+4}}{1+(1-2\epsilon_{\pi/8})^{4N+4}} =(6 + 14 N + 8 N^2)\epsilon_{\pi/8}^2  + O(\epsilon_{\pi/8}^3)
\end{equation}
This suffices to conclude that  
\begin{equation}
\epsilon' \leq (2N-1) \epsilon_{\theta}^2 + (6 + 14 N + 8 N^2)\epsilon_{\pi/8}^2  + N \eta + O(\epsilon_{\theta}^3, \epsilon_{\pi/8}^3, \eta^2, \ldots) 
\end{equation}
where $O(\epsilon^3)$ collects all higher order terms.  For instance, for $N=1$ and $N=2$ this yields
\begin{align}
	\epsilon' & \leq \begin{cases}   \epsilon_{\theta}^2 + 28 \epsilon_{\pi/8}^2  + \eta + O(\epsilon_{\theta}^3, \epsilon_{\pi/8}^3, \eta^2, \ldots) & \mbox{ for } N=1 \\
	3\epsilon_{\theta}^2 + 66 \epsilon_{\pi/8}^2  + 2 \eta + O(\epsilon_{\theta}^3, \epsilon_{\pi/8}^3, \eta^2, \ldots) & \mbox{ for } N=2 \\
	\end{cases}
\end{align}
Comparing this with the numerical expressions Eq.~\eqref{Numerical_error_out} and Eq.~\eqref{Numerical_error_out2}, we see the analytic upper bound is very loose and grossly overestimates the prefactors. 

\section{No small triorthogonal codes}
\label{Sec_No_tri}

Many other distillation protocols are based upon projections into codespaces with a transversal non-Clifford, with transversality proofs typically using some notion of triorthogonal matrices~\cite{Bravyi12}.  While the protocols proposed here do not manifestly have this form, it is natural to ask whether there is some codespace projection with equivalent performance.   Indeed, Jones' first-level distiller protocol~\cite{Jones13} is effectively equivalent to projecting onto the codespace of Bravyi-Haah triorthogonal codes~\cite{Bravyi12}, and Haah has recently introduced level-lifting as a general methodology for finding such equivalences~\cite{haah17towers}.  Furthermore, it has long been known that for any distillation protocol there exists a codespace projection that achieves the same error suppression~\cite{Camp09c}, though it may not achieve the same success probability or admit a transversal non-Clifford gate. 

In this section, we show that there exist no triorthogonal codes with fewer than 14 qubits.   This bound is tight since the smallest Bravyi-Haah code is a 14 qubit triorthogonal construction.  It follows that the  $10 \rightarrow 2$ MEK protocol is not equivalent to a projection onto a triorthogonal code.  What distinguishes MEK is that it is a highly compressed circuit that is obtained from taking a larger circuit and cancelling some $T$-gates.  This suggests that something happens during the compression process of eliminating extraneous $T$-gates that breaks the equivalence to triorthogonal codes.  Since our protocols can be understood as a generalisation of MEK protocols, it seems unlikely similar performance parameters will be achievable using projections onto codes with exotic transversality properties.

We present the definition of triorthogonality
\begin{defin}
	(Def 1. of Ref. \cite{Bravyi12})A binary matrix $G$ of size $m \times n$ is called triorthogonal iff the supports of any pair and any triple of its rows have even overlap, that is,
	\begin{equation}
		\sum_{j=1}^n G_{a,j}G_{b,j} = 0  \pmod{2}
	\end{equation}
	for all pairs of rows $1 \leq a < b \leq m$ and
   \begin{equation}
	\sum_{j=1}^n G_{a,j}G_{b,j}G_{c,j} = 0  \pmod{2}
	\end{equation}
	for all triples of rows $1 \leq a < b < c \leq m$.
\end{defin}	
The definition of triorthogonality allows a matrix to have either odd or even rows, and it is standard to use a horizontal line to demarcate the split
\begin{equation}
G = \left( \frac{G_1}{G_0} \right),
\end{equation}
so $G_1$ contains odd weight rows and $G_0$ contains even weight rows.  Assuming $G$ is row-wise linearly independent, it describes an $[[n,k,d]]$ quantum code where: $n$ is the number of columns in $G$; $k$ is the number of rows in $G_1$; and $d \geq 2$ if and only if $G_0$ is non-trivially supported on every column.  We also use the notion of a biorthogonal matrix, which obeys the constraint for pairs of rows but not for triples of rows.

Let $G$ be a triorthogonal matrix with block matrix form
\begin{equation}
G = \left( \frac{G_1}{G_0} \right) = \left( \begin{array}{cc}
A & B \\ \hline
C & D \\
\mathbf{1} 	&	\mathbf{0} 
\end{array}
\right) ,
\end{equation}
where 	$\mathbf{1} $ and $\mathbf{0}$ are the all-1 and all-0 row vectors of appropriate width.  Using column permutation the matrix can always be brought into this form.   Without loss of generality, we assume that the last row has weight $w$ and is the highest weight row in the span of $G_0$.  Let $B$ and $D$ be width $u$, so that the total matrix width is $n=w+u$.

From triorthogonality of $G$, it follows that the submatrix
\begin{equation}
L=	\left( \begin{array}{c} A \\ C \end{array}  \right) ,
\end{equation}
is biorthogonal with all rows being even weight and that 
\begin{equation}
R=	\left( \begin{array}{c} B \\ \hline D \end{array}  \right) ,
\end{equation}
is biorthogonal with $B$ containing odd weight rows.   Since $B$ contains odd weight rows, the matrix $D$ cannot contain the all-1 vector as this would violate biorthogonality.  However, since the code is distance 2, the matrix $D$ must be supported on every column.  Therefore, there must exist at least 2 non-trivial rows in $D$.   The smallest possible width for $D$ is then achieved by
\begin{equation}
D = \left( \begin{array}{cccccc} 1 & 1 & 1 & 1 & 0 & 0  \\ 1 & 1 & 0 & 0 & 1 & 1  \end{array}  \right),
\end{equation}	
which has width $u=6$ and contains weight 4 vectors, and so we can infer that $w \geq 4$.  Other $D$ are possible, but one cannot obtain smaller parameters:  There are at least two rows and for any pair of rows they must overlap on at least 2 columns and also have support on at least 2 other non-overlapping columns. From this we see that $n = w + u \geq 4+6 = 10$.  So this is already enough to prove there are no $[[n,k,2]]$ triorthogonal codes with $k \leq 1$ and $n$ less than 10.   

However, the bound $w \geq 4$ was obtained based on the rows of $D$ only, but $w$ is the max weight across all rows in the span of $G_0$.  If $C$ contains a row of weight $w_c$, then we know that $w \geq w_c + 4$.  For any row of $C$ we can add the last row of $G_0$, which generates a row of weight $w_c' =w - w_c$, entailing that $w \geq w_c' + 4 = w - w_c + 4$ and so $w_c \geq 4$.  Putting this together yields $w \geq 8$ and $u \geq 6$ so that $n \geq 14$.  There are no triorthogonal codes with fewer than 14 qubits.

\section{Discussion}

\subsection{Variation of the two-step protocol}

This subsection discusses one possible variant of the two-step protocol.  Consider a quantum algorithm that needs many magic states of the form $\ket{R(\theta)}$, but with different values of $\theta$. As presented, our main protocol cannot be used to full effect as the very large $N$ limit assumes that we need many magic states with the same $\theta$.  However, it is straightforward to check that one can distill pairs of states with the same $\theta_j$.  That is, we may input states of the form $\bigotimes_{j=1}^N \rho(\theta_j)^{\otimes 2}$ and use $N$ pivotal rotations with corresponding angles $2\theta_j$. 

\subsection{Quadratic vs higher order error suppression}
\label{sec_quadratic}

In our introduction, we remarked on the existence of distillation routines that offer much larger reductions in errors without using concatenation~\cite{Jones13,haah17magic,haah2017examples,hastings2017sublog}.  The appeal of these protocols is better asymptotic performance in the limit of large quantum computers and large error reduction.  The analysis underpinning these results assumes that it is appropriate to quantify resource costs by the ratio of input to output states.   But a more realistic picture is given by an involved full space time analysis; also accounting for the cost of Clifford gates and quantum error correction.  In such an analysis, it is possible to scale the size of error correction codes between rounds of magic state distillation~\cite{Raussendorf06,Fowler12,gorman17}.  This scaling trick is extremely effective, and is arguably the most important tool in the arsenal of magic state distillation techniques.  Although the idea has been known for some time, it has gone without a name.  In an effort to popularise this trick, O'Gorman and Campbell recently proposed the phrase ``balanced investment''~\cite{gorman17}.

Balanced investment relies on distinct rounds of magic state distillation with successive error reduction.  Therefore, balanced investment is more compatible with protocols giving quadratic error reduction, such as presented here, than with the protocols of Refs.~\cite{Jones13,haah17magic}.  This argument is qualitative, and we need detailed resource analyses to make concrete quantitative statements.  However, such full resource investigations are difficult and time-consuming and have only been undertaken for the Reed-Muller and Bravyi-Haah protocols.  Naturally, such a numerical investigation also falls outside the scope of this paper.

\section{Conclusions}

We presented a new two-step method of magic state distillation that is very competitive at preparing single-qubit magic states, offering a way to circumvent the need for costly gate-synthesis of single-qubit rotations.  An important aspect of these new protocols are the preparation of multi-qubit magic states using synthillation. Pressing open questions include how these competing approach fare when all resource costs are considered, though such an analysis will depend heavily on the architecture considered.   

We would also like to explore whether the synthillation driven techniques proposed here could be extended to protocols with larger than quadratic error reduction~\cite{Jones13,haah17magic}.   After completing this work, Hastings and Haah proposed some new approaches to synthillation that may provide a starting point for attacking this problem~\cite{haah17synth}.

\section{Acknowledgements}

This research was supported by the EPSRC (grant ref EP/M024261/1).  We thank IBM and developers of the QISKit, which was used for numerical simulations.

 \bibliographystyle{unsrtnat}
% \bibliography{MagicLib5}

\begin{thebibliography}{28}
	\providecommand{\natexlab}[1]{#1}
	\providecommand{\url}[1]{\texttt{#1}}
	\expandafter\ifx\csname urlstyle\endcsname\relax
	\providecommand{\doi}[1]{doi: #1}\else
	\providecommand{\doi}{doi: \begingroup \urlstyle{rm}\Url}\fi
	
	\bibitem[Campbell et~al.(2017)Campbell, Terhal, and Vuillot]{ReviewPaper}
	Earl~T Campbell, Barbara~M Terhal, and Christophe Vuillot.
	\newblock Roads towards fault-tolerant universal quantum computation.
	\newblock \emph{Nature}, 549\penalty0 (7671):\penalty0 172, 2017.
	\newblock \doi{10.1038/nature23460}.
	
	\bibitem[Bravyi and Kitaev(2005)]{BraKit05}
	Sergey Bravyi and Alexei Kitaev.
	\newblock Universal quantum computation with ideal {C}lifford gates and noisy
	ancillas.
	\newblock \emph{Phys. Rev. A}, 71:\penalty0 022316, 2005.
	\newblock \doi{10.1103/PhysRevA.71.022316}.
	
	\bibitem[Meier et~al.(2013)Meier, Eastin, and Knill]{Meier13}
	Adam~M. Meier, Bryan Eastin, and Emanuel Knill.
	\newblock Magic-state distillation with the four-qubit code.
	\newblock \emph{Quant. Inf. and Comp.}, 13:\penalty0 195, 2013.
	
	\bibitem[Bravyi and Haah(2012)]{Bravyi12}
	Sergey Bravyi and Jeongwan Haah.
	\newblock Magic-state distillation with low overhead.
	\newblock \emph{Phys. Rev. A}, 86:\penalty0 052329, Nov 2012.
	\newblock \doi{10.1103/PhysRevA.86.052329}.
	
	\bibitem[Jones(2013{\natexlab{a}})]{Jones13}
	Cody Jones.
	\newblock Multilevel distillation of magic states for quantum computing.
	\newblock \emph{Phys. Rev. A}, 87:\penalty0 042305, Apr 2013{\natexlab{a}}.
	\newblock \doi{10.1103/PhysRevA.87.042305}.
	
	\bibitem[Haah et~al.(2017)Haah, Hastings, Poulin, and Wecker]{haah17magic}
	Jeongwan Haah, Matthew~B. Hastings, D.~Poulin, and D.~Wecker.
	\newblock Magic state distillation with low space overhead and optimal
	asymptotic input count.
	\newblock \emph{{Quantum}}, 1:\penalty0 31, October 2017.
	\newblock ISSN 2521-327X.
	\newblock \doi{10.22331/q-2017-10-03-31}.
	
	\bibitem[Kliuchnikov et~al.(2013)Kliuchnikov, Maslov, and Mosca]{kliuchnikov13}
	Vadym Kliuchnikov, Dmitri Maslov, and Michele Mosca.
	\newblock Asymptotically optimal approximation of single qubit unitaries by
	clifford and $t$ circuits using a constant number of ancillary qubits.
	\newblock \emph{Phys. Rev. Lett.}, 110:\penalty0 190502, May 2013.
	\newblock \doi{10.1103/PhysRevLett.110.190502}.
	
	\bibitem[Gosset et~al.(2014)Gosset, Kliuchnikov, Mosca, and Russo]{gosset14}
	David Gosset, Vadym Kliuchnikov, Michele Mosca, and Vincent Russo.
	\newblock An algorithm for the t-count.
	\newblock \emph{Quant. Inf. \& Comp.}, 14\penalty0 (15-16):\penalty0
	1261--1276, 2014.
	
	\bibitem[Ross and Selinger(2016)]{RS14}
	Neil~J Ross and Peter Selinger.
	\newblock Optimal ancilla-free clifford+ t approximation of z-rotations.
	\newblock \emph{Quant. Inf. and Comp.}, 16:\penalty0 901, 2016.
	
	\bibitem[Paetznick and Svore(2014)]{paetznick14}
	Adam Paetznick and Krysta~M Svore.
	\newblock Repeat-until-success: Non-deterministic decomposition of single-qubit
	unitaries.
	\newblock \emph{Quant. Inf. \& Comp.}, 14\penalty0 (15-16):\penalty0
	1277--1301, 2014.
	
	\bibitem[Bocharov et~al.(2015)Bocharov, Roetteler, and Svore]{bocharov15}
	Alex Bocharov, Martin Roetteler, and Krysta~M. Svore.
	\newblock Efficient synthesis of probabilistic quantum circuits with fallback.
	\newblock \emph{Phys. Rev. A}, 91:\penalty0 052317, May 2015.
	\newblock \doi{10.1103/PhysRevA.91.052317}.
	
	\bibitem[Landahl and Cesare(2013)]{landahl13}
	Andrew~J Landahl and Chris Cesare.
	\newblock Complex instruction set computing architecture for performing
	accurate quantum $ z $ rotations with less magic.
	\newblock \emph{arXiv preprint arXiv:1302.3240}, 2013.
	\newblock URL \url{https://arxiv.org/pdf/1302.3240.pdf}.
	
	\bibitem[Jones(2013{\natexlab{b}})]{jones13fourier}
	Cody Jones.
	\newblock Distillation protocols for fourier states in quantum computing.
	\newblock \emph{arXiv preprint arXiv:1303.3066}, 2013{\natexlab{b}}.
	\newblock URL \url{https://arxiv.org/pdf/1303.3066.pdf}.
	
	\bibitem[Duclos-Cianci and Poulin(2015)]{duclos15}
	Guillaume Duclos-Cianci and David Poulin.
	\newblock Reducing the quantum-computing overhead with complex gate
	distillation.
	\newblock \emph{Phys. Rev. A}, 91:\penalty0 042315, Apr 2015.
	\newblock \doi{10.1103/PhysRevA.91.042315}.
	
	\bibitem[Campbell and O'Gorman(2016)]{smallAngle}
	Earl~T Campbell and Joe O'Gorman.
	\newblock An efficient magic state approach to small angle rotations.
	\newblock \emph{Quantum Science and Technology}, 1\penalty0 (1):\penalty0
	015007, 2016.
	\newblock \doi{doi:10.1088/2058-9565/1/1/015007}.
	
	\bibitem[Haah(2017)]{haah17towers}
	Jeongwan Haah.
	\newblock Towers of generalized divisible quantum codes.
	\newblock \emph{arXiv preprint arXiv:1709.08658}, 2017.
	\newblock URL \url{https://arxiv.org/pdf/1709.08658.pdf}.
	
	\bibitem[Jones(2013{\natexlab{c}})]{jones13b}
	Cody Jones.
	\newblock Low-overhead constructions for the fault-tolerant toffoli gate.
	\newblock \emph{Phys. Rev. A}, 87:\penalty0 022328, Feb 2013{\natexlab{c}}.
	\newblock \doi{10.1103/PhysRevA.87.022328}.
	
	\bibitem[Eastin(2013)]{eastin13}
	Bryan Eastin.
	\newblock Distilling one-qubit magic states into toffoli states.
	\newblock \emph{Phys. Rev. A}, 87:\penalty0 032321, Mar 2013.
	\newblock \doi{10.1103/PhysRevA.87.032321}.
	
	\bibitem[Campbell and Howard(2017{\natexlab{a}})]{campbell17}
	Earl~T. Campbell and Mark Howard.
	\newblock Unifying gate synthesis and magic state distillation.
	\newblock \emph{Phys. Rev. Lett.}, 118:\penalty0 060501, Feb
	2017{\natexlab{a}}.
	\newblock \doi{10.1103/PhysRevLett.118.060501}.
	
	\bibitem[Campbell and Howard(2017{\natexlab{b}})]{campbell17b}
	Earl~T. Campbell and Mark Howard.
	\newblock Unified framework for magic state distillation and multiqubit gate
	synthesis with reduced resource cost.
	\newblock \emph{Phys. Rev. A}, 95:\penalty0 022316, Feb 2017{\natexlab{b}}.
	\newblock \doi{10.1103/PhysRevA.95.022316}.
	
	\bibitem[Haah et~al.(2018)Haah, Hastings, Poulin, and Wecker]{haah2017examples}
	Jeongwan Haah, Matthew~B Hastings, D~Poulin, and D~Wecker.
	\newblock Magic state distillation at intermediate size.
	\newblock \emph{Quant. Inf. and Comp.}, 18:\penalty0 0114, 2018.
	
	\bibitem[Hastings and Haah(2018)]{hastings2017sublog}
	Matthew~B. Hastings and Jeongwan Haah.
	\newblock Distillation with sublogarithmic overhead.
	\newblock \emph{Phys. Rev. Lett.}, 120:\penalty0 050504, Jan 2018.
	\newblock \doi{10.1103/PhysRevLett.120.050504}.
	
	\bibitem[Gottesman and Chuang(1999)]{CliffHier}
	Daniel Gottesman and Isaac~L. Chuang.
	\newblock Demonstrating the viability of universal quantum computation using
	teleportation and single-qubit operations.
	\newblock \emph{Nature}, 402:\penalty0 390, 1999.
	\newblock \doi{10.1038/46503}.
	
	\bibitem[Campbell and Browne(2009)]{Camp09c}
	Earl~T. Campbell and Dan~E. Browne.
	\newblock On the structure of protocols for magic state distillation.
	\newblock \emph{Lecture Notes in Computer Science}, 5906:\penalty0 20, 2009.
	\newblock \doi{10.1007/978-3-642-10698-9_3}.
	\newblock arXiv:0908.0838.
	
	\bibitem[Raussendorf et~al.(2006)Raussendorf, Harrington, and
	Goyal]{Raussendorf06}
	R.~Raussendorf, J.~Harrington, and K.~Goyal.
	\newblock A fault-tolerant one-way quantum computer.
	\newblock \emph{Annals of Physics}, 321\penalty0 (9):\penalty0 2242 -- 2270,
	2006.
	\newblock ISSN 0003-4916.
	\newblock \doi{10.1016/j.aop.2006.01.012}.
	
	\bibitem[Fowler et~al.(2012)Fowler, Mariantoni, Martinis, and
	Cleland]{Fowler12}
	Austin~G. Fowler, Matteo Mariantoni, John~M. Martinis, and Andrew~N. Cleland.
	\newblock Surface codes: Towards practical large-scale quantum computation.
	\newblock \emph{Phys. Rev. A}, 86:\penalty0 032324, Sep 2012.
	\newblock \doi{10.1103/PhysRevA.86.032324}.
	
	\bibitem[O'Gorman and Campbell(2017)]{gorman17}
	Joe O'Gorman and Earl~T. Campbell.
	\newblock Quantum computation with realistic magic-state factories.
	\newblock \emph{Phys. Rev. A}, 95:\penalty0 032338, Mar 2017.
	\newblock \doi{10.1103/PhysRevA.95.032338}.
	
	\bibitem[Haah and Hastings(2017)]{haah17synth}
	Jeongwan Haah and Matthew~B Hastings.
	\newblock Codes and protocols for distilling $ t $, controlled-$ s $, and
	toffoli gates.
	\newblock \emph{arXiv preprint arXiv:1709.02832}, 2017.
	\newblock URL \url{https://arxiv.org/pdf/1709.02832.pdf}.
	
\end{thebibliography}

\end{document}